\documentclass[prd,twocolumn,showpacs,letterpaper,nofootinbib]{revtex4}
\usepackage{graphics}
\usepackage{epsfig} 
\usepackage{amsmath}
\input epsf

\newcommand{\be}{\begin{equation}}
\newcommand{\ee}{\end{equation}}
\newcommand{\bea}{\begin{eqnarray}}
\newcommand{\eea}{\end{eqnarray}}
\def\pslash{{\cal P}{\hbox{\kern-6pt $\slash$}}}

\long\def\comment#1{}
\begin{document}
%\draft
%\twocolumn[\hsize\textwidth\columnwidth\hsize\csname @twocolumnfalse\endcsname
\title{Dark-energy dependent test of general relativity at cosmological scales.}
\author{Yves Zolnierowski}
\email{zolniero@lapp.in2p3.fr}
\affiliation{Laboratoire d'Annecy-le-Vieux de
Physique des Particules, CNRS/IN2P3 and Universit\'e Savoie Mont Blanc, 9 Chemin de Bellevue, BP 110,
F-74941 Annecy-le-Vieux c\'edex, France}
\author{Alain Blanchard}
\email{alain.blanchard@irap.omp.eu}
\affiliation{Universit\'e de Toulouse, UPS-OMP, CNRS, IRAP,  F-31028 Toulouse,  France}

\date{\today}

\begin{abstract}

The $\Lambda$CDM framework offers a remarkably good description of our universe with a very small number of free parameters, 
which can be determined with high accuracy from currently available data. 
However, this does not mean that the associated physical quantities, such as the curvature of the universe, have been directly measured. 
Similarly, 
general relativity is assumed, but not tested. 
Testing the relevance of  general relativity for cosmology at the background level includes a verification of the relation 
between its energy contents and the curvature of space. 
Using an extended Newtonian formulation, we propose an approach where 
 this relation can be tested.  
Using the recent measurements on cosmic microwave background,  baryonic acoustic oscillations and the supernova Hubble diagram, 
we show that the prediction of general relativity is well verified 
in the framework of standard $\Lambda$CDM assumptions, i.e. an energy content only composed  
of matter and dark energy, in the form of a cosmological constant or equivalently a vacuum contribution. 

However, the actual equation of state of dark fluids cannot be directly obtained from cosmological observations. We found that   
relaxing the  equation of state of dark energy opens a large region of possibilities, 
revealing a new type of degeneracy between the curvature and the total energy content of the universe.

\end{abstract}

%]
\maketitle
\vskip 1cm

\section{INTRODUCTION}

 The  near-Euclidean nature of the universe  was one of the main predictions of 
inflationary cosmology \cite{Guth}. Indeed inflationary models actually predict that the curvature radius of the  
universe should be much larger than the Hubble scale $ct_0$ 
($t_0$ being the present age of the universe), provided enough e-fold occurred.  This 
leads to a tight prediction on the spatial-curvature parameter $|\Omega_k|<10^{-4}$ \cite{Linde2008} 
 which can only be bypassed at the price of unnatural fine-tuning \cite{BullKamion}. 
 Within the $\Lambda$CDM framework, 
the curvature of our universe can be constrained from the observed cosmic microwave background (CMB) fluctuations \cite{LB1998} and  
 current estimations of curvature parameter are achieving impressive accuracy. 
For instance Planck collaboration \cite{Planck15XIII} reported $\Omega_k = 0.0008\pm 0.0004 (95\%)$ 
when the CMB  and  some other cosmological data are combined. These recent results 
appear as a remarkable evidence in favor of inflationary cosmology \cite{PlanckXXII},   while the  value of the  curvature radius of the universe  
appears as an important and fundamental characteristics of the universe.
The successful   standard $\Lambda$CDM picture  is  leading to 
the existence of dominant and unknown dark components: the non-baryonic dark matter  and the dark energy, 
which is used as the generic term for the origin of the accelerated expansion of the universe,
as revealed by the Hubble diagram of supernovae type Ia \cite{Riess98,Perlmutter}. 
While there is almost no doubt on the reality of the acceleration of the expansion of the universe, 
 its physical origin eludes us.
 A large variety of explanations have been proposed:  
scalar field domination known as quintessence,  generalized gravity theory beyond  general relativity (GR) or 
inhomogeneous cosmological models. 
The present day situation is therefore quite paradoxical:   within the simplest framework of GR 
and $\Lambda$CDM picture, cosmological parameters can be determined with a few percent accuracy but on the other hand  
we have little evidence on the validity of this framework. 
It is certainly desirable 
to test the theory at the foundation of our cosmological 
model \cite{CK2004}. The remarkable  consistency of the present universe with a spatially flat space and 
its importance for the theory of inflation, 
 both encourage for a close examination of the actual constraints that can be set and to their connections 
with the energy contents of the universe.
In the standard analyses aimed at measuring  cosmological parameters, the approach used does not offer a measurement  
of the  curvature of space, as would provide a purely geometrical test \cite{Bernstein2006}. 
 As a crucial property of our universe,  one would like to measure the spatial geometry of our space in a model-independent way. 
The knowledge of the luminosity distance with redshift does not allow to determine 
 spatial curvature of our 3D space \cite{Weinberg1970}, and should be complemented by the measurement 
of another observational quantity such as the Hubble  expansion rate history $H(z)$.  
 This can be obtained from
longitudinal BAO \cite{EuclidTh} or 
from the look-back time of the universe \cite{JL2002,MJ2011}. 
With the latter method, a first estimate of the spatial curvature leads to a result consistent with flatness but with uncertainties of 
the order of unity \cite{MJ2011}. 
As GR implies that the 
curvature of the universe is related to its energy content, an independent measurement of 
the latter 
would allow 
to test if
GR holds at cosmological scales. Given the remarkable accuracy of GR at  describing gravity at small scales, 
it is clearly important to check whether GR  actually holds on cosmological scales. The relation 
between the curvature and energy contents is precisely such a test of GR at the background level. 
In the following, we focus on how to examine the status of this relation given the quality of relevant cosmological data. \\

\section{FROM NEWTONIAN GRAVITY TO GENERAL RELATIVITY}

The commonly used metric of a space-time in which 3D spatial slices are homogeneous  is the Robertson-Walker metric:
\begin{equation}
d s^2 = - c^2d t^2 +a(t)^2\left( {{d r^2}\over {1-kr^2}}+ r^2(d\theta^2+\sin^2\theta d\phi^2) 
\right) 
\end{equation}
in which $a(t)$ is the expansion factor of the universe and $k$ is a constant related to the geometry of space.
 A value of $k = +1$ corresponds to a spherical space, $k = 0$ to an euclidean space and $k = -1$ to a hyperbolic space. \\
The knowledge of the $a(t)$ function  allows to derive the coordinate $r$:
\begin{equation}
r  =  S_{k} \left(\int_{t_{\rm S}}^{t_0}{{c dt }\over{a(t)}} \right)=  S_{k} \left(\int^{z_{\rm S}}_{0}{{c dz }\over{H(z)}} \right)
\label{eq:r}
\end{equation}
where:
\begin{equation}
   {S_k(X)} \ = \ \left\{ \begin{array}{l}
\sin (X) \\ 
 X \\  \sinh (X) \end{array}\right. \ \text{ for } \ \begin{array}{lcl}
 k & = &  +1 \\ 
 k & = & 0  \\  
 k & = &  -1
\end{array}
\end{equation}
and thereby the angular-diameter distance:
\begin{equation}
D_A(z) = a(t)r = \frac{a_0r}{1+z}
\label{eq:da}
\end{equation}
or equivalently the luminosity distance ($D_L(z) = D_A(z)(1+z)^2$).
Eq. \ref{eq:da} makes it clear that the knowledge of  $D_A(z)$ is not sufficient to infer both the curvature $a_0$ 
and the dynamical evolution of the expansion $a(t)$ \cite{Weinberg1970}. 
This leads to a well-known degeneracy between the curvature 
and a possible evolving dark energy \cite{Knox2006, Clarkson2007, Virey2008}. 
Breaking these
degeneracies within GR is possible by using different observables,
 such as for instance growth rate measurements \cite{Mortonson2009}. 
 Indeed, the dynamic of perturbations may provide  information which cannot be obtained from the evolution of 
the background  \cite{Linder2005,Bertchi2006}, allowing to distinguish modified gravity from dark energy models\cite{B&Z2008}.

The dynamical evolution of the expansion factor $a(t)$ is directly related to the energy content 
of the universe and to the underlying theory of gravity. 
It is possible to derive the equation governing $a(t)$ in Newtonian cosmology, by writing the mechanical energy of a sphere, 
 the radius of which can be taken arbitrarily small (the full justification relies on the Birkhoff-Jebsen's theorem). 
Because the equation does not depend on the radius, GR should lead exactly to 
the same equation \cite{Mukhanov2005}. This equation reads:
\begin{equation}
\left(\frac{\dot{a}}{a}\right)^2 + \frac{K}{a^2} = \frac{8\pi G}{3} \sum \rho_i
\label{eq:dynnew}
\end{equation}
where the $\rho_i$ are the energy densities of the different fluids contributing to the gravitational field 
of the universe and $K$ an arbitrary constant. 

 The normalized curvature parameter $\Omega_k=-kc^2/(Ha)^2$, where $H $ is the Hubble parameter $\dot{a}/a$, 
is introduced to quantify the geometry of the universe. In a similar way, we can define 
 a dynamical quantity:
\begin{equation}
\Omega_{k_{dyn}}=1-\sum \Omega^c_i\;
\end{equation}   
where $\Omega^c_i$ are the density parameters of the various energy contents of the universe 
($\displaystyle \Omega^c_i =  8 \pi G \rho_i / (3 H^2)$). 
In a theory leading to Newtonian dynamic for $a(t)$ (eq. \ref{eq:dynnew}), the relation would be of the form:
\begin{equation}
\Omega_{k_{dyn}} = \alpha  \Omega_{k_{geo}}
\end{equation}   
with $\alpha = K/(kc^2)$ being a  constant of the theory.

When derived in the frame of GR, the constant $K$ is not arbitrary anymore but related to the  geometric constant 
$k$ found in the Robertson-Walker metric:
\begin{equation}
K=kc^2\;
\end{equation}   
i.e. $\alpha = 1$, leading to the well-known  Friedman-Lema\^{\i}tre equation which now
reads:
\begin{equation}
\Omega_k = 1 - \sum \Omega^c_i
\end{equation}

\section{A SIGNATURE OF GENERAL RELATIVITY IN COSMOLOGY}

Alternative theories of gravity to GR offer a potential origin 
for  the acceleration of the expansion, and have been thoroughly investigated in recent years. 
 As we have seen that independent measurements of $\Omega_k$ 
on one side and of the $\Omega^c_i$ on the other side allow to test the validity of GR.
 This test also holds for 
other theories which would exhibit a  different relation between curvature and 
matter-energy content such as the backreaction models \cite{Larena2009} or some $f(R)$ models. 
 It is clear that  these two quantities are physically independent and the  prediction of GR is:
\begin{equation}
\Omega_{k_{geo}} = \Omega_{k_{dyn}}.
\end{equation}   

The Hubble expansion factor can now be written as:
\begin{eqnarray}
H(z)^2 &=& H_0^2 E(z)^2 \label{eq:Hz}\\
 &=& H_0^2 \left[\Omega_m (1 + z)^3 + \Omega_{k_{dyn}} (1 + z)^2 + \Omega_{DE}(z)\right].\nonumber
\end{eqnarray}

 An estimate of curvature of space has been recently obtained  \cite{Okouma2013} from distance measurements, 
independently   of assumptions about the evolution of  dark energy, but still within the context of GR. 
As we mentioned in the beginning, methods exist for  measuring the 
curvature of our 3D space  independently of  the underlying theory of gravity 
by using distances and measurements of the expansion rate $H(z)$, allowing to test the 
Copernican principle by estimating $\Omega_k(z)$ \cite{Clarkson2008}. 
$H(z)$ could be inferred from the  evolution  of redshift with time \cite{Zshift,Corasetal2007},  
by differentiating the age of the universe \cite{JL2002} or by Baryon Acoustic Oscillation
(BAO) measurements \cite{Clarkson2007}. 

 The situation is more problematic when one wants to estimate $\sum \Omega^c_i$ from the observations. 
In GR  the energy-momentum tensor can be evaluated from the 
 measurement of the local quantities $\rho(t)$ and $p(t)$ and their properties. 
However, the dark energy fluids are likely to escape to local measurements and their properties 
can probably be constrained only through observations at cosmological scales.

Indeed, a combination of geometrical observations (at the background level) 
cannot provide  a way to disentangle  the total energy content  in distinct components such as pressureless  matter
and dark energy on the other side.
In the absence of local measurements of the content of the universe, densities and pressure, we are left with the option
  to parametrize the observables \cite{Kunz2009}.
 
 In the following study, we therefore adopt a parametrized description of the dynamical contents 
of the universe and examine whether the prediction of GR is verified. 
 
Dark energy fluids are often characterized by their equation of state $w(t)$ relating pressure and density: $ p(t) = w(t) \rho(t)$. 
The simplest dark energy model is certainly a cosmological constant or equivalently a vacuum contribution, 
corresponding to $w(t) = -1$. Therefore, we  investigate this simple 
 scenario where  matter and vacuum make up the contents of the universe 
and then examine the case of a dark energy component with a constant $w$.

\section{DATA SETS AND METHODS USED}

The constraints on the two new parameters of the model, $\Omega_{k_{geo}}$ and $\Omega_{k_{dyn}}$, 
are obtained by a minimization of total
likelihood
$\chi^2$ with the help of a custom Markov chain program. 
We start with eight random points in the parameter space and test 
the convergence of the chains with the Gelman-Rubin diagnostics \cite{1992GelmanRubin} 
 which is  based on the comparison of the variance of the parameters
 inside and between the chains. When convergence is established, the parameters of each step are stored.
 For each cell in the $\Omega_{k_{geo}}-\Omega_{k_{dyn}}$ parameter space, our program computes the number of 
chain steps and the minimum $\chi^2$ inside it.
 
The three standard cosmological observables, SN1a,
large surveys of galaxies and CMB fluctuations are used by the program.\\
The supernovae set is the JLA sample composed of 740 SN1a resulting from a combination of the first three years of SNLS and 
the three seasons of the SDSS-II  completed with 14 high $z$ SN1a from HST and several low $z$ samples of SNIa \cite{Betoule2014}. 
For this work only the diagonal terms of the error matrix are  used.
The expression of the luminosity distance, $D_L$ is modified to introduce the new parameters:
\begin{eqnarray}
D_L  =   \cfrac{c \ (1+z)}{H_0 \ \sqrt{|\Omega_{k_{geo}}}|} \ {S_k} \left( \sqrt{|\Omega_{k_{geo}}|} \int_0^z \cfrac{ \ {d}u}{E(u)}\right) 
\end{eqnarray}
The BAO resulting from the couplings between the gravitational forces 
and electromagnetic forces in the primordial plasma produce a specific pattern in both (the fluctuating part of) 
the radiation field and the matter density field. These features 
provide a powerful tool for  constraining the cosmological parameters.
The constraints coming from BAO and CMB fluctuations are established  
with  the help of reduced parameters 
published by different collaborations and authors. In practice we  
 use the $\chi^2_{BAO}$ presented 
in the final publication of the WMAP collaboration \cite{wmap9}.
This $\chi^2_{BAO}$ is built using the values of the quantity $r_s(z_d)/D_V(z_{mes})$ or $D_V(z_{mes})/r_s(z_d)$ 
at different redshifts $z_{mes}$ published by 6dFGS \cite{2011Beutler}, SDSS-DR7-rec \cite{2012Padmanabhan},  
SDSS-DR9-rec\cite{2012Anderson} and WiggleZ \cite{2011Blake}.
The quantity $D_V(z_{mes})$, combination of radial and transverse distances is defined by the relation  
$D_V(z_{mes}) = \left[c z (1+z_{mes})^2 D_A^2(z_{mes}) / H(z_{mes}) \right]^{1/3}$ 
at redshift $z_{mes}$ where
$D_A$ is the angular-diameter distance between today and redshift $z$ and 
$r_s(z_d)$, the physical sound horizon at the end of the drag area (redshift $z_d$).
$r_s(z_d)$ is given 
by the relation $r_s(z_d) = \int_{z_d}^\infty c_s(u)/H(u) {d}u$ with $c_s(u)$ the sound speed in the primordial plasma.
The sound speed at redshift $z$ is related to the ratio of the baryon density, $\rho_b$, over photon  density $\rho_\gamma$, 
$R_b(z)$ by the relation 
$c_s(z) = c / \sqrt{3 (1 + R_b(z))}$ where $R_b(z) = 3 \rho_b / (4 \rho_\gamma)$.

The value of $r_s$ is computed 
in our approach using the expression \ref{eq:Hz} of $H(z)$  
with the addition of an extra term for the contribution of the relativistic particles at the time of radiation matter equality.\\

The statistical properties of the CMB fluctuations which are a powerful source of constraints on cosmological parameters, 
can be summarized through 
the following  three quantities:
the shift parameter, $R(z_{*}) = \sqrt{\Omega_m H_0^2} (1 + z_{*}) D_A(z_{*}) / c$, 
the acoustic scale, $l_A(z_{*}) = \pi (1 + z_{*}) D_A(z_{*}) / r_s(z_{*})$, and the baryon density  $\Omega_b h^2$ 
where $z_{*}$ is the redshift of the decoupling between matter and radiation.
We use the values of these 3 parameters and the associated covariance matrix provided by 
\cite{Wang2013} using the Planck archive data. \\
The fitting formulas of  Hu-Sugiyama  and Eisenstein-Hu are used to compute $z_d$ et $z_{*}$ \cite{1996HuSugi,1999E&Hu}.
The use of these fitting formulas is justified by the fact that these scales are determined by the physics of 
the matter-baryons-photon fluids, 
well understood in terms of atomic physics and linear perturbations \cite{Vonlanthen2010}. 
These fitting formulas fully apply as long as the dark energy fluid contribution remains negligible prior to the drag period $z \geq z_d$.  
\begin{figure}[tb]
\centerline{
\includegraphics[width=0.95\columnwidth]{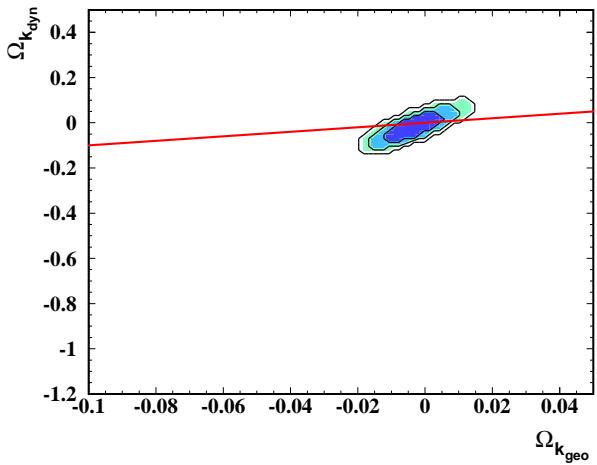}}
\caption{1, 2 and 3 $\sigma$ contours in the  $\Omega_{k_{geo}} - \Omega_{k_{dyn}}$ plane  
for models with a dark energy component $w = -1$. The line corresponds to the prediction of GR. 
In the Hubble expansion factor $H(z)$, the $\Omega_m$ parameter is tightly constrained by the Planck shift parameter $R(z_{*})$.}
\label{fig:omk1omk2}
\end{figure}

\begin{figure}[tb]
\centerline{
\includegraphics[width=0.95\columnwidth]{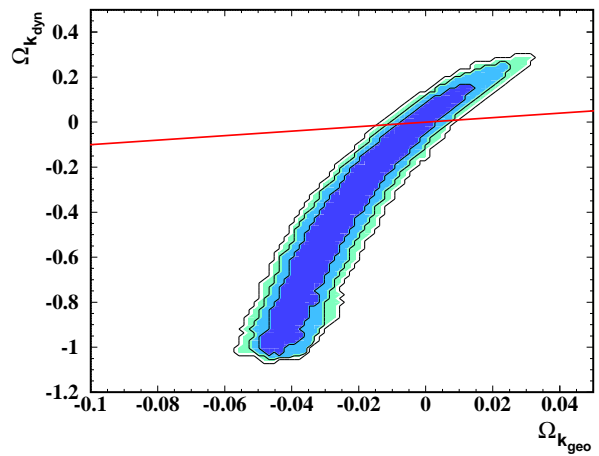}}
\centerline{
\includegraphics[width=0.95\columnwidth]{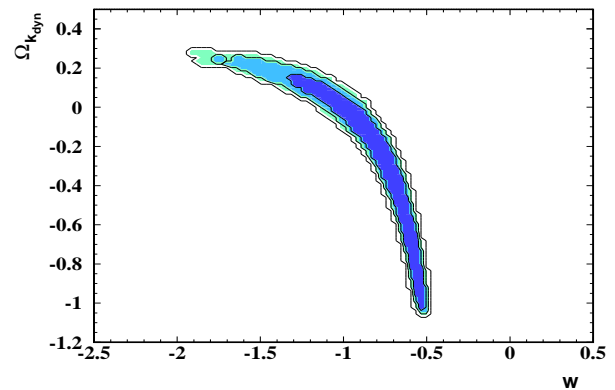}}
\caption{ 1, 2 and 3 $\sigma$ contours in the  $\Omega_{k_{geo}} - \Omega_{k_{dyn}}$
y for models having a dark energy component with an arbitrary constant equation of state parameter $w$. 
The contours  are marginalized for $ -2.5 < w < 0.5 $. 
The addition of a new degree of freedom, $w$, 
 widens the contours revealing a degeneracy between $\Omega_{k_{dyn}}$ and $\Omega_{k_{geo}}$}
\label{fig:omk1omk2w}
\end{figure}

\section{DISCUSSION AND CONCLUSIONS}

We have established a  simple approach to test GR on cosmological scales at the  background level. 
This approach provides a new type of constraints on the curvature of space without 
assuming GR (but requiring assumptions on dark energy evolution) and separates clearly 
the geometrical and the dynamical curvatures.  
Using the three standard probes, we found interesting constraints  when assuming a dark energy component 
in the form of a vacuum energy ($w = -1$):
\begin{equation}
\nonumber
\left\{ \begin{array}{lcl} 
 \Omega_{k_{geo}} & = & -0.0058\pm 0.0052 \ (1 \sigma) \\
 \Omega_{k_{dyn}} & = & -0.043\pm 0.034  \ (1 \sigma)
\end{array} \right.
\end{equation}
 Our estimate on $\Omega_{k_{geo}}$ and more importantly its uncertainties,   are close to the ones 
obtained by the Planck collaboration for $\Omega_k$, while the uncertainties on  our parameter $\Omega_{k_{dyn}}$ are noticeably worse. 

At this level, however, from our analysis we found that GR is entirely consistent 
with existing data with little room for any alternative. This offers a new evidence for the remarkable agreement 
 between the standard  GR+$\Lambda$CDM model in view of and modern data in Cosmology.  It is quite remarkable that introducing 
a new degree of freedom (i.e. one parameter extension of the $\Lambda$CDM model) 
does not change the basic figures on the properties of our universe in its simplest version.\\ 
The situation changes drastically when the assumption on dark energy is relaxed. 
 For example, allowing for an additional  dark energy fluid with $w = -1/3$ in the contents of the universe will produce 
a degeneracy with the actual curvature and will not allow to test directly GR at the background level with this approach. \\
With an arbitrary constant equation of state parameter $w$ we found: 
\begin{equation}
\nonumber
\left\{ \begin{array}{lcl} 
 \Omega_{k_{geo}} & = & -0.025\pm 0.03  \ (1 \sigma) \\
\nonumber
 \Omega_{k_{dyn}} & = & -0.45\pm 0.6  \ (1 \sigma)\\
 w  & = & -0.75^{+0.25}_{-0.55}  \ (1 \sigma)
\end{array} \right.
\end{equation}
 This considerably enlarges the possibilities with a significant room 
for alternative theories to standard GR: there is a significant degeneracy between curvature and energy contents. 
This may appear somehow not surprising as the determination of the energy contents suffers from a degeneracy problem 
that has been discussed several times in the past. 
However, the situation is quite different from the standard model as the same additional freedom on dark energy $w$ does not 
modify much the constraints. For instance from WMAP7, BAO and SN one gets
 $ w \sim -1.0 \pm 0.06$ and $\Omega_{k} = -0.006\pm 0.007$ \cite{wmap7cc}.
  On the other hand, the limits on $\Omega_{k_{geo}}$  that we obtained are similar to those obtained in GR 
based models without any assumption  
on dark energy \cite{Okouma2013} and similar to those based on model-independent methods \cite{Lietal2014,Sapone2014,Heavens2014}. 
While similar, our limits on $\Omega_{k_{geo}}$ 
clearly depends on our assumptions on dark energy: constant equation of state  $w$ 
(and the speed of sound equal to unity). The conclusion is quite interesting: 
relaxing the assumption on dark energy leads to much wider allowed range for $\Omega_{k_{dyn}}$ 
with an equation of state for dark energy which  is not well constrained, $-1.3 < w < -0.5$ (1 $\sigma$) 
and with an energy content which 
is poorly known: $-1.105 \leq \Omega_{k_{dyn}} \leq 0.15$ (1 $\sigma$). 
It is reasonable to think that more freedom on dark energy 
will lead to even looser constraints, a question needing further investigation . \\
It is quite remarkable that the standard picture (GR+$\Lambda$CDM) fits so well the present day data
while at the same time we do not have enough independent data yet to clarify the nature of dark energy and 
therefore the appropriateness of the  theory of GR. The true universe may well be much more complicated. 
This provides further motivation for a deeper investigation of the  very nature of dark energy.\\ \\

\textbf{Acknowledgements.}  We thank St\'ephane Ilic, Martin Kunz and Brahim Lamine for useful comments. Yves Zolnierowski acknowledges financial support
provided by IRAP (UMR5277/CNRS/UPS).


\begin{thebibliography}{99}
  

\bibitem{Guth}   A.~H. Guth, \prd {\bf 23}, 347 (1981).

\bibitem{Linde2008} A. Linde, Lect. Notes Phys.   {\bf 738}, 1 (2008).


\bibitem{BullKamion} P. Bull  and M. Kamionkowski,  \prd {\bf 87}, 081301 (2013).

\bibitem{LB1998} C.~H. Lineweaver and D. Barbosa,  Astrophys. J.  {\bf 496}, 624  (1988).

\bibitem{Planck15XIII} Planck 
Collaboration, P.~A.~R. Ade {\it et al.}, arXiv:1502.1589.

\bibitem{PlanckXXII}   Planck 
Collaboration, P.~A.~R. Ade {\it et al.},  Astron. Astrophys., {\bf 571}, AA22 (2014).

\bibitem{Riess98} A.~G., Riess  {\it et al.},
 Astron. J. {\bf   116}, 1009 (1998).

\bibitem{Perlmutter}S. Perlmutter  {\it et al.}, Astrophys. J. {\bf 517}, 565 (1999).

\bibitem{CK2004} R.~R. Caldwell and  M. Kamionkowski, 
J. Cosmol.
Astropart. Phys.  09 (2004) 09.

\bibitem{Bernstein2006} G. Bernstein, Astrophys. J.
 {\bf 637}, 598 (2006).

\bibitem{Weinberg1970} S. Weinberg,   Astrophys. J. Letters
 {\bf  161}, L233  (1970).

\bibitem{EuclidTh} L. Amendola  {\it et al.}, Living Reviews in Relativity {\bf 16}, 6 (2013).

\bibitem{JL2002} R. Jimenez and A. Loeb,   Astrophys. J. {\bf 573}, 37 (2002).

\bibitem{MJ2011} E. Mortsell and J. Jonsson, arXiv:1102.4485.

\bibitem{Clarkson2007} C. Clarkson,  
M. Cort{\^e}s, and B. Bassett,  J. Cosmol.
Astropart. Phys.  { 08}  (2007)  11.

\bibitem{Knox2006} L. Knox,  \prd {\bf 73}, 023503  (2006).

\bibitem{Virey2008} J.-M. Virey {\it et al.}, J. Cosmol. Astropart. Phys.   12 (2008) 008.

\bibitem{Mortonson2009} M.~J. Mortonson, 
\prd {\bf 80}, 123504 (2009).

\bibitem{Linder2005} E.~V. Linder,   \prd {\bf  72}, 
043529 (2005).

\bibitem{Bertchi2006} E. Bertschinger,  Astrophys. J. 
 {\bf  648}, 797  (2006).

\bibitem{B&Z2008} E. Bertschinger and P. Zukin \prd {\bf 78}, 024015 (2008).

\bibitem{Mukhanov2005} V. Mukhanov,  Physical 
Foundations of Cosmology. Cambridge 
University Press,  (2005).  


\bibitem{Kunz2009} M. Kunz,   \prd {\bf 80}, 123001 (2009).

\bibitem{Larena2009} J. Larena, J.-M. Alimi, T. Buchert, M. Kunz, P.S. Corasaniti,  \prd {\bf 79}, 083011 (2009).


\bibitem{Okouma2013} P.~M., Okouma,  Y. Fantaye, 
 and  B.~A. Bassett,  Phys. Letters B  {\bf  719}, 1 (2013).

\bibitem{Clarkson2008} C. Clarkson, B. Bassett and Lu TeresaHui-Ching, \prl {\bf 101}, 011301 (2008).

\bibitem{Zshift} A. Sandage,  Astrophys. J. {\bf  136}, 319 (1962).

\bibitem{Corasetal2007} P.-S. Corasaniti, D. Huterer and A. Melchiorri,  \prd {\bf 75}, 062001 (2007). 

\bibitem{1992GelmanRubin} A. Gelman and D. B. Rubin, Statist. Sci. {\bf 7},  457-511 (1992).

\bibitem{Betoule2014} M. Betoule {\it et al.}, Astron. Astrophys.  {\bf 568}, A22 (2014).
 
\bibitem{wmap9} G. Hinshaw {\it  et al.}, Astrophys. J.  Suppl. Ser. {\bf  208}, 19 (2013).

\bibitem{2011Beutler} F. Beutler {\it et al.}, Mon. Not. R. Astron. Soc. {\bf  416}, 3017 (2011).  

\bibitem{2012Padmanabhan} N. Padmanabhan {\it et al.}, Mon. Not. R. Astron. Soc. {\bf 427}, 2132 (2012).

\bibitem{2012Anderson} L. Anderson {\it et al.}, Mon. Not. R. Astron. Soc. {\bf 427}, 3435  (2012).

\bibitem{2011Blake} C. Blake {\it  et al.}, Mon. Not. R. Astron. Soc., 418, 1725 (2011). Mon. Not. R. Astron. Soc., {\bf 425}, 405 (2012).

\bibitem{Wang2013} Y. Wang, S. Wang, \prd {\bf 88}, 043522  (2013).

\bibitem{1996HuSugi} W. Hu and N.  Sugiyama,  Astrophys. J. {\bf  471}, 542 (1996).

\bibitem{1999E&Hu} D.~J. Eisenstein and  W. Hu,    Astrophys. J. {\bf  511}, 5 (1999).

\bibitem{Vonlanthen2010} M. Vonlanthen, S. R\"as\"anen and R. Durrer, 
J. Cosmol. Astropart. Phys.   08 (2010) 023.

\bibitem{wmap7cc} E., Komatsu
 
{\it  et al.}, Astrophys. J. Suppl. Ser. {\bf 192}, 18 (2011).


\bibitem{Lietal2014} Y.-L. Li, S.-Y.  Li, T.-J. Zhang, 
and  T.-P. Li,  Astrophys. J. Letters {\bf 789}, L15 (2014).

\bibitem{Sapone2014} D. Sapone, E. Majerotto 
and S.  Nesseris,   \prd {\bf 790}, 023012 (2014).

\bibitem{Heavens2014} A. Heavens,  R. Jimenez and
 L. Verde, \prl {\bf 113}, 241302  (2014).


\end{thebibliography}
\end{document}